\documentstyle[preprint,prd,aps,psfig]{revtex}\tighten

\newcommand{\beq}{\begin{equation}}
\newcommand{\eeq}{\end{equation}}
\newcommand{\beqa}{\begin{eqnarray}}
\newcommand{\eeqa}{\end{eqnarray}}
\newcommand{\hr}{\widehat{r}}
\newcommand{\hphi}{\widehat{\phi}}
\newcommand{\hv}{\widehat{V}}
\newcommand{\k}{\kappa}
\begin{document}
\widetext
\draft
\preprint{\tighten\vbox{\hbox{KUNS-1677}\hbox{YITP-00-41}
\hbox{astro-ph/0008175}}}

\title{
Feasibility of  Reconstructing the Quintessential Potential\\ 
Using SNIa Data 
}

\author{Takeshi Chiba}

\address{
Department of Physics, Kyoto University,
Kyoto 606-8502, Japan}

\author{Takashi Nakamura}
\address{
Yukawa Institute for Theoretical Physics,
Kyoto University,
Kyoto 606-8502, Japan}

\date{August 10, 2000}

\maketitle

\bigskip

\begin{abstract}
We investigate the feasibility of the method for reconstructing the
equation of state and the effective potential of the quintessence field 
from SNIa data. We introduce a useful functional form to fit the
luminosity distance with good accuracy (the relative error is less than
0.1\%). We assess the ambiguity in reconstructing the equation of state
and the effective potential which originates from the uncertainty in 
$\Omega_M$. We find that the equation of state is sensitive to the
assumed $\Omega_M$, while the shape of the effective potential is not. 
We also demonstrate the actual reconstruction procedure using the data
created by Monte-Carlo simulation. Future high precision measurements of 
distances to thousands of SNIa could reveal the shape of the
quintessential potential.

\end{abstract}



\section{Introduction}
Recent various observations \cite{data,data2}, in particular distance
measurements to SNIa \cite{sn1,sn2}, strongly suggest that the universe 
is currently dominated by a positive vacuum energy density with negative 
pressure. The smallness of the vacuum energy density 
$\sim (10^{-12}{\rm GeV})^4$ has 
revived the idea that the cosmological ``constant'' is not really a
constant but rather decaying. The idea of quintessence \cite{cds} (see
also \cite{csn,tw} and references therein) is that vacuum
energy density is played by a scalar field rolling down the almost flat
potential similar to cosmological inflation, and a lot of models have
been proposed so far. However, there is currently no clear guidance
from particle physics as to which quintessence models may be suitable. 
Then it should be the observations that decide which model is correct
or not. As  a bottom-up approach, we have proposed that distance 
measurements to SNIa may allow one to reconstruct the equation of state
of the dark energy or the effective potential of the quintessence 
field \cite{nc}.

Future observational project, such as SNAP(SuperNova/Acceleration
Probe)\footnote{http://snap.lbl.gov} could gather 2,000 SNIa in a single 
year and could put significant constraint on the cosmological parameters 
(including the equation of state of the dark energy $w$). In view of the 
future prospect for high-z SNIa search, we investigate in detail the
feasibility of the method for reconstructing the equation of state and
the effective potential of the quintessence field from SNIa data
\cite{nc,recon,alex}. \footnote{While our work was being completed, we 
became aware of related work \cite{st} which focuses on the uncertainty
in the reconstruction of the equation of state of dark energy.}

\section{Parameterizing the luminosity distance}

Considering the future prospect for high-z SNIa search, we believe it 
particularly useful to fit the observed luminosity distance $d_{L}(z)$
to a function of $z$ which has the following properties: (1)the good
convergence (the relative error is hopefully less than 0.1\% because the 
distance error expected from SNAP will be less than  a few percent)
for $0< z < 10$; (2) the correct asymptotic behavior for  $z \gg 1$
($H(z) \propto (1+z)^{3/2}$). We present a fitting function for
$d_{L}(z)$.

We restrict ourselves to a flat FRW universe henceforth and assume
Einstein gravity.\footnote{The reconstruction equation for the so called 
extended quintessence \cite{extq} is presented in \cite{es}.}   
In a flat model, the luminosity distance $d_{L}(z)$ is written
in terms of the coordinate distance $r(z)$ to an object at $z$ as
\beq
{d_{L}(z)\over (1+z)}=r(z)=\int^{t_0}_{t}{dt'\over
  a(t')}=\int^{z}_{0}{dz'\over H(z')}=\int^1_y{2dy'\over H(y')y'^3},
\eeq
where $t_0$ is the present time and $y=1/\sqrt{1+z}$. It is interesting to 
note that in terms of $y$, $r(y)$ is a linear function of $y$ 
if $\Omega_{M}=1$. Therefore, the $d^2r(y)/dy^2$ contains the information 
of the non-zero pressure (see Eq.(\ref{reconst:eos}) below). 
We will elaborate on the advantage of using $y$ and $r$ in a separate
paper. 

In analogy with Pen's powerful fitting formula for $r(z)$ for a flat
FRW universe with a cosmological constant \cite{pen}, we propose to fit
$r(z)$ in the following functional form\footnote{An extension to an open 
or closed model is immediate once $\Omega_K$ is known: 
$H_0r(z)=|\Omega_k|^{-1/2}\sin_K\left(|\Omega_k|^{1/2}
(\eta(1)-\eta(y))\right)$, where $\sin_K(x)=\sin(x) (\sinh(x))$ if
$K=1(-1)$.}:\beqa
&&H_{0}r(z)=\eta(1)-\eta(y)\\
&&\eta(y)=2\alpha\left[y^{-8}+\beta y^{-6}+\gamma
  y^{-4}+\delta y^{-2}+\sigma\right]^{-1/8}.
\label{fit}
\eeqa
The requirement $H(z)/H_{0} \rightarrow 1$ for $z
\rightarrow 0$ imply that $\sigma$ is found to be a dependent parameter: 
$\sigma=\left(\alpha\left(1+3\beta/4+\gamma/2+\delta/4\right)\right)^{8/9}-
1-\beta-\gamma-\delta$.
However, we shall treat $\sigma$ as a free parameter for simplicity in
numerical calculations. 
For $y \rightarrow 0~ (z \rightarrow \infty)$, we have
$H(z)/H_{0}\rightarrow 1/(\alpha y^{3})$. For the Einstein-de Sitter
universe, $\alpha=1,\beta=\gamma=\delta=0$. 

\subsection{demonstrating the goodness of fit}

We shall demonstrate that the fitting function Eq.(\ref{fit}) 
 indeed does a good job. For this purpose, we calculate the maximum 
relative error in $r(y)$ between the actual value and that calculated
{} from the fitting function Eq.(\ref{fit}) in the  range $0.3 < y < 1~
 (10 > z> 0)$. We consider cosmological models consisting of matter and
dark energy of constant equation of state $w\equiv
 p_X/\rho_X=0,-1/3,-2/3,-1$ with $0.1\leq \Omega_M\leq 1$.  
We fit each template luminosity distance, $r_i$, by the functional form 
Eq.(\ref{fit}) using the Davidon-Fletcher-Powell method. 
We minimize $\sum_{y_i}(r_i-r(y_i))^2/r_i^2$ with N=30 data points. 
The result is shown in Fig. 1. We find the maximum relative error is
less than 0.05\%. 
Our fitting function thus seems more powerful by order of magnitude than 
the one proposed in \cite{alex}.

\begin{figure}[htdp]
  \begin{center}
  \leavevmode\psfig{figure=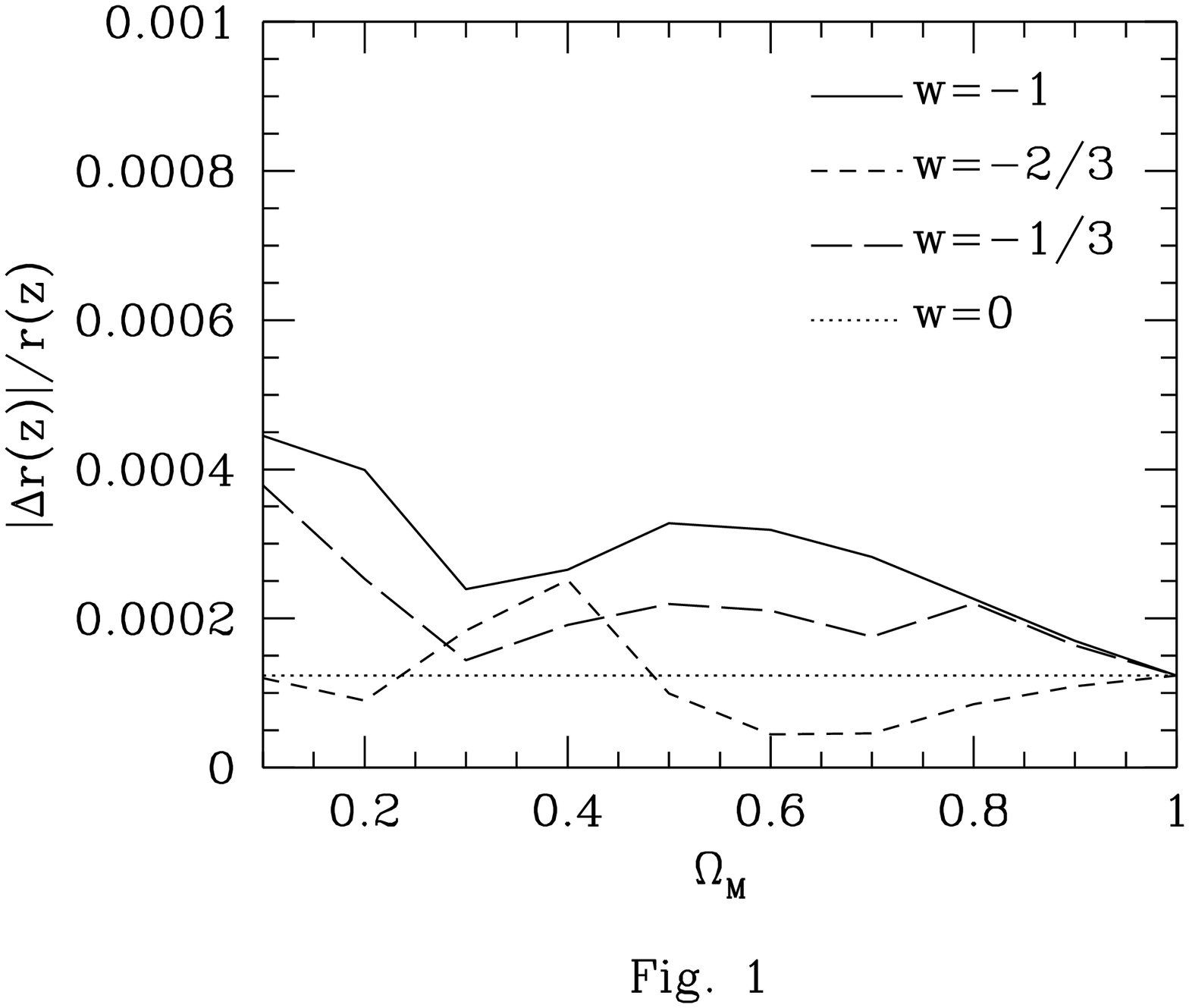,width=9cm}
  \end{center}
  \caption{The maximum deviation in the range 
     $0.3 < y < 1~ (10 > z> 0)$ between the actual value and that
     calculated from our fitting function is shown as a function of 
       $\Omega_M$.}
  \label{fig:fig1}
\end{figure}

\section{Reconstructing the Equation of State and the Effective Potential}

One of the prime interest in the reconstruction issue is whether the
effective equation of state of the x-component 
$w\equiv p_X/\rho_X$
is different from $-1$, which is the unique signature of the dynamical 
vacuum energy. 
If observations would suggest $w\neq -1$, then the next
urgent project would be the real reconstruction of the effective
potential, which should have profound implications for particle 
physics as well as for cosmology. 

In this section, we reconstruct the equation of state of dark energy and 
the effective potential of the quintessence field. 
Regarding the energy density of the Universe, we assume two components:
nonrelativistic  matter with its present density parameter $\Omega_{M}$
and x-component with $\Omega_X=1-\Omega_{M}$. An extension to 
include the radiation component is immediate but with negligible effect.

\subsection{reconstructing the equation of state}

In terms of a dimensionless variable $\hr\equiv H_{0}r$, the
 equation-of-state of x-component $w$ is written as
\beqa
w(z)&=&{3d\hr(z)/dz+2(1+z)d^{2}\hr(z)/dz^{2}
\over 3\left(d\hr(z)/dz\right)\left(\Omega_{M}
\left(d\hr(z)/dz\right)^{2}(1+z)^{3}-1\right)}\\
&=&{-4y d^{2}\hr/dy^{2}
\over 3\left(d\hr/dy\right)\left(\Omega_{M}
\left(d\hr/dy\right)^{2}-4\right)}.
\label{reconst:eos}
\eeqa
$w$ thus depends on the second derivative of the luminosity
function. Put another way, the luminosity distance depends on $w$ 
through a multiple integral relation \cite{st}.
Whether $w=-1$ or not will clearly signify whether the dark 
energy is constant in time or not. 

\begin{figure}[htdp]
  \begin{center}
  \leavevmode\psfig{figure=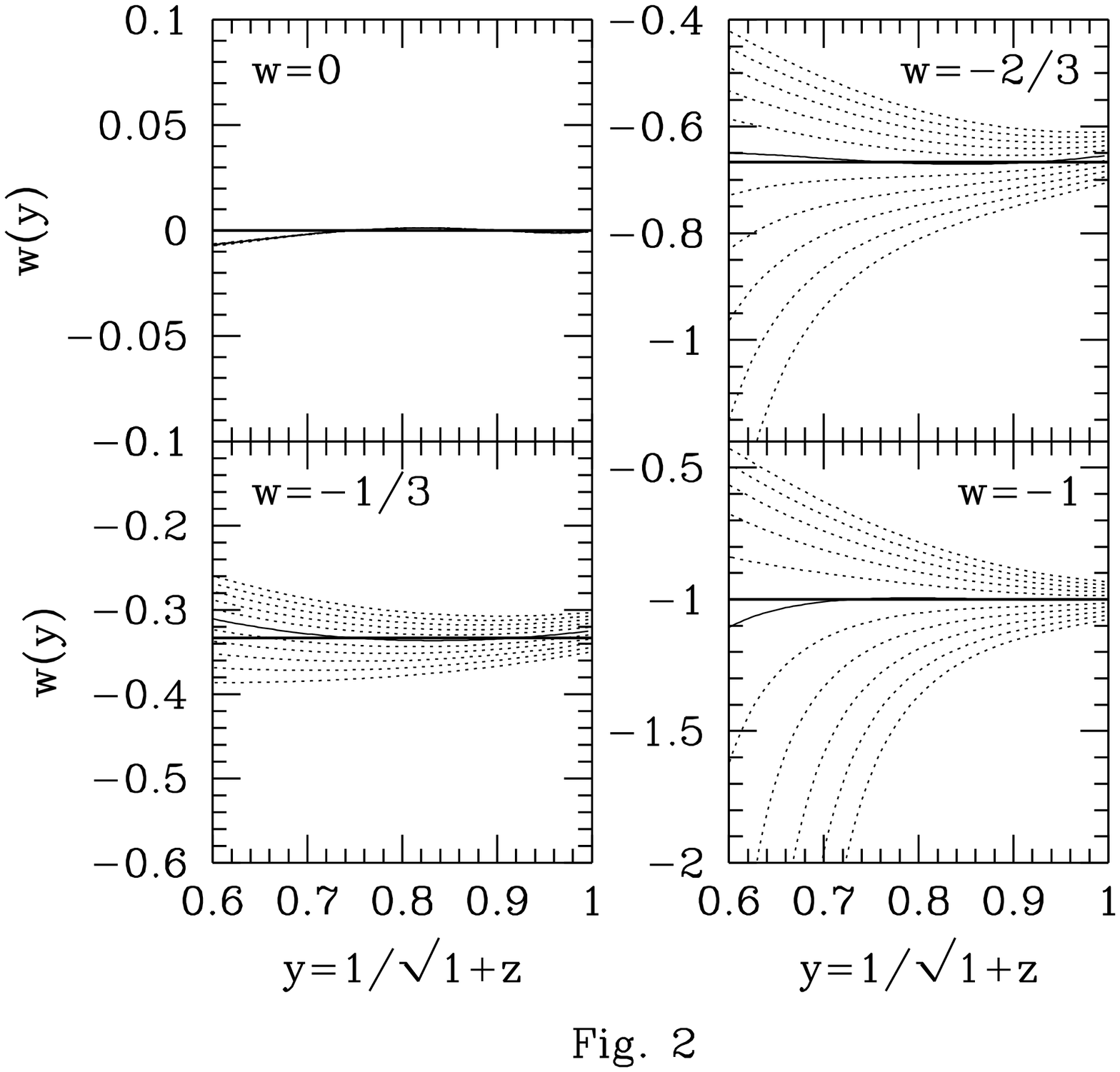,width=12.5cm}
  \end{center}
  \caption{The reconstructed equation of state assuming $\Omega_M=0.30$ 
(solid curves). The curves assuming 
$\Omega_M=0.25,0.26, \dots,0.35$ (from top to bottom) are shown by
dotted ones.}
  \label{fig:fig2}
\end{figure}

In Fig. 2, we show the reconstructed $w(y)$ for cosmological models 
with constant equation of state $w=0,-1/3,-2/3,-1$ to examine the effect
of the ambiguity in $\Omega_M$.
The template $\hr_i$ is constructed by 
assuming $\Omega_M=0.30$ for $0.60<y<1~(1.78>z>0)$. The template is
fitted using the ansatz Eq.(\ref{fit}) by minimizing 
$\sum_{y_i}(\hr_i-\hr(y_i))^2/\hr_i^2$ with N=10 data points. 
We also plot the reconstructed $w(y)$ for
$\Omega_M=0.25,0.26,\dots,0.35$(from top to bottom) as dotted curves. 
We find that the uncertainty in $\Omega_M$ would enlarge an error in 
$w(y)$ for small $y$ (large $z$). In particular, when we overestimate 
$\Omega_M$, $w(y)$ may diverge at some $y$ where the denominator in 
Eq.(\ref{reconst:eos}) may vanish. However, an error in $w$ remains
relatively small near $y=1$: 10\% uncertainty in $\Omega_M$ results in
at most 19\% error in $w$  for $y> 0.80~(z<0.56)$. The error is largest
for the cosmological constant (19\% error) and becomes less significant 
for larger $w$. For example, for $w=-2/3$ model, the error is less than
12\% for $y> 0.80$. Hence it might be possible to discriminate between
$w=-1$ and $w\neq -1$. The combination of SNIa measurements and high
precision measurements of the power spectrum expected from the Microwave
Anisotropy Probe (MAP) and Planck satellites could make a clear 
distinction \cite{ptw}.

\subsection{reconstructing the effective potential}

Reconstructing equations are 
\beqa
\hv(z)&=&{3\over
  \left(d\hr(z)/dz\right)^{2}}+(1+z){d^{2}\hr(z)/dz^{2}\over
  \left(dr(z)/dz\right)^{3}}-{3\over 2}\Omega_{M}(1+z)^{3},\\
\left({d\hphi (z)\over
    dz}\right)^{2}&=&{\left(d\hr(z)/dz\right)^{2}\over (1+z)^{2}}
\left[-2(1+z){d^{2}\hr(z)/dz^{2}\over \left(d\hr(z)/dz\right)^{3}}
-3\Omega_{M}(1+z)^{3}\right],
\eeqa
where $\hphi\equiv \k\phi, \hv\equiv \k^{2}V/H_{0}^{2}$ with
$\k^{2}=8\pi G$. Alternatively, in terms of $y=1/\sqrt{1+z}$
\beqa
\hv(y)&=& {6\over y^6(d\hr/dy)^2}-{2d^2\hr/dy^2\over
  y^5(d\hr/dy)^3}-{3\Omega_M\over 2y^6},\label{reconst:v}\\
\left({d\hphi (y)\over dy}\right)^{2}&=&
{12\over y^2} +{4 d^2\hr/dy^2\over y (d\hr/dy)}-3\Omega_M
{ (d\hr/dy)^2\over y^2}.
\label{reconst:p}
\eeqa

In order to demonstrate the effectiveness of the fitting function
Eq.(\ref{fit}), we reconstruct the effective potential of the
quintessence field. We consider three kinds of potentials which have
some theoretical backgrounds; (a) cosine
type \cite{cosine}, $\hv(\hphi)=M^4(\cos(\hphi)+1)$; (b) inverse
power law type 
\cite{power}, $\hv(\hphi)=M^4\hphi^{-\alpha}$; (c) exponential type
\cite{exp}, $\hv(\hphi)=M^4\exp(-\lambda \hphi)$. 
$M$ and $\lambda$ are fixed to  give $\Omega_X=1-\Omega_M$. 
$\alpha=4$ is assumed hereafter. 

The results are shown in Fig. 3. The dotted curves are numerically
reconstructed $w(y)$ and $\hv(\hphi)$ with $\Omega_M=0.30$ being assumed, 
while the solid curves are the original ones up to $y=0.6$. 
We fix the present value of $\hphi$ to unity. 

We also plot the dashed curves assuming $\Omega_M=0.25,0.27,0.33,0.35$.
Since the smaller $y$, the larger the error in $w(y)$, the range of
$\hphi$ significantly depends on the assumed $\Omega_M$. We note that
for $\Omega_M>0.30$, $w$ becomes less than $-1$ and thus 
the right hand side of Eq.(\ref{reconst:p}) turns negative at some $y$. 
So we stop the reconstruction of $\hv(\hphi)$ there. 
The range of $\hphi$ is larger (smaller) for smaller (larger)
$\Omega_M$. For example, in the case of cosine potential, the true range
of $\hphi$ is $0.41<\hphi \leq 1.00$, while the reconstructed range is 
$0.41<\hphi \leq 1.00$ for $\Omega_M=0.30$,  $0.17<\hphi \leq 1.00$ for 
$\Omega_M=0.25$, $0.25<\hphi \leq 1.00$ for $\Omega_M=0.27$, 
$0.59<\hphi \leq 1.00$ for $\Omega_M=0.33$, and $0.66<\hphi \leq 
1.00$ for $\Omega_M=0.35$. 
However, it is interesting that the whole shape of $\hv(\hphi)$ is less
sensitive to the uncertainty in $\Omega_M$, although $w$ and the range 
of $\hphi$ are dependent on the assumed value of $\Omega_M$. 
If we assume smaller $\Omega_M$, the resulting potential energy is larger 
via Eq.(\ref{reconst:v}). On the other hand, $\hphi$ decreases more rapidly 
back in time via Eq.(\ref{reconst:p}). The opposite is the case for 
larger $\Omega_M$. Both effects make the reconstructed 
shape of $\hv(\hphi)$ converge to the true one.

\begin{figure}[htdp]
  \begin{center}
  \leavevmode\psfig{figure=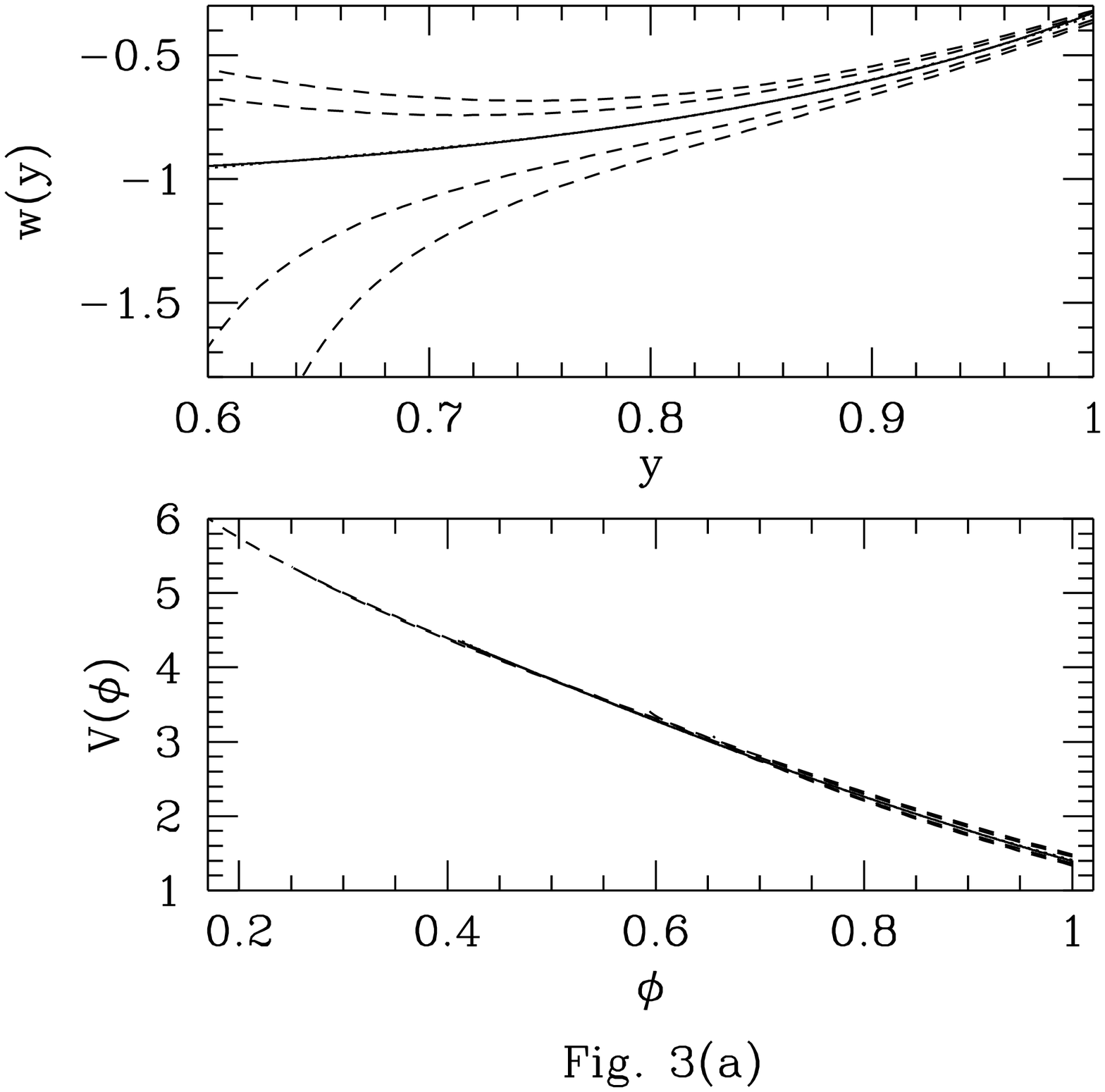,width=8cm}
  \leavevmode\psfig{figure=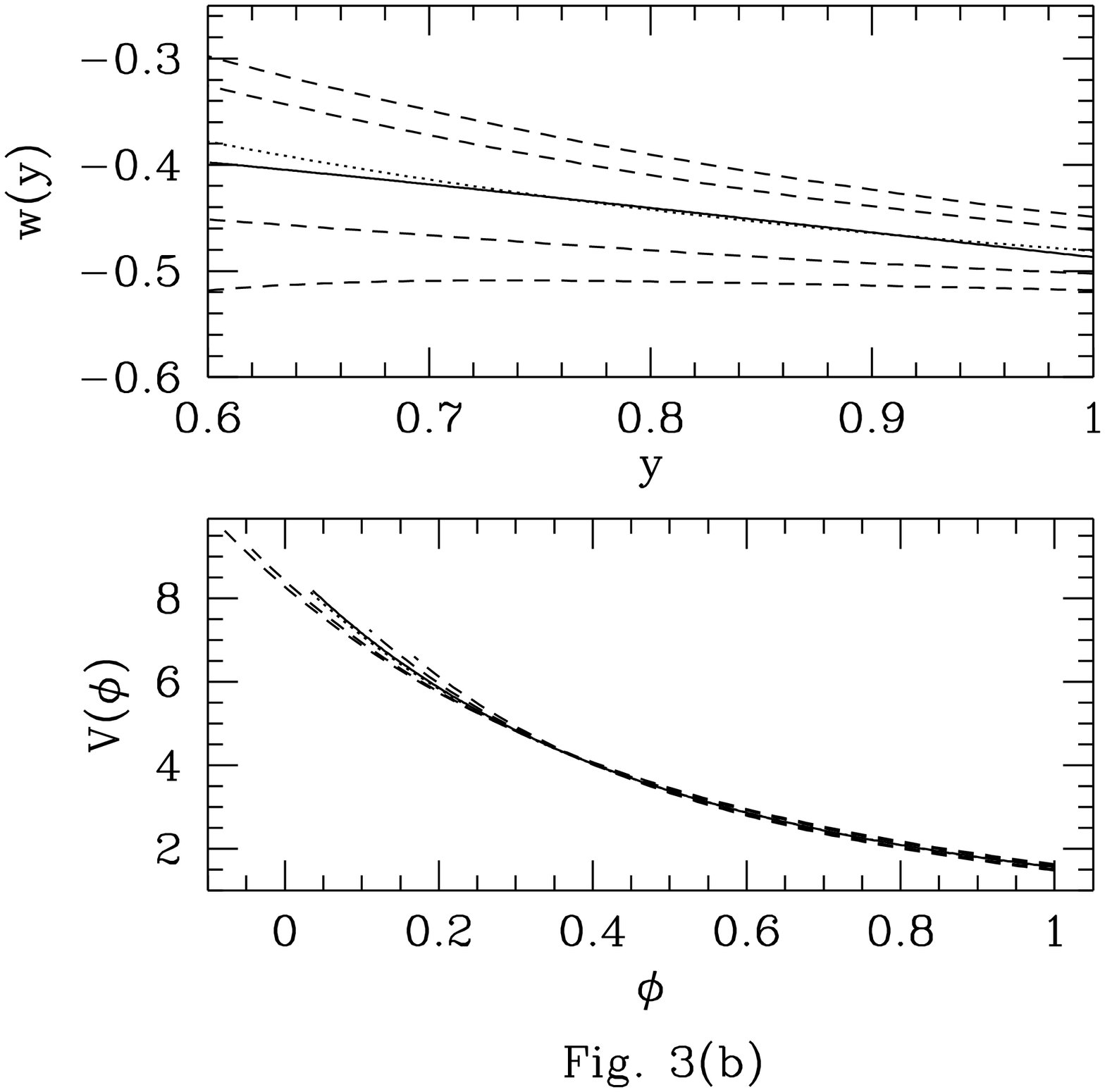,width=8cm}
  \leavevmode\psfig{figure=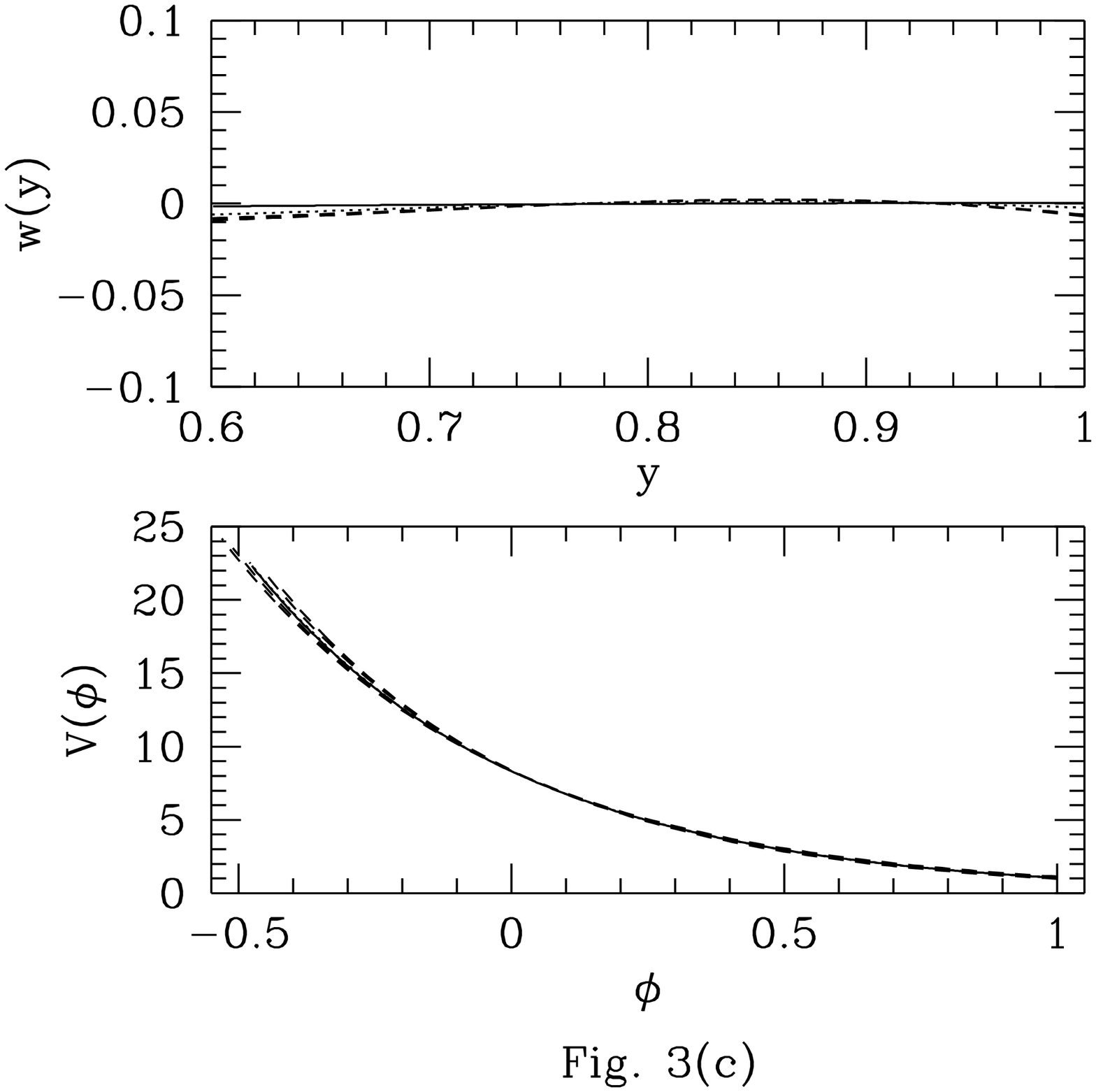,width=8cm}
  \end{center}
  \caption{The reconstructed equation of state and the effective
  potential: (a) cosine potential, (b) inverse power law potential, (c)
  exponential potential. Solid curves are the original equations of
  state or effective potentials. Dotted curves are for $\Omega_M=0.30$. 
  The curves assuming $\Omega_M=0.25,0.27,0.33,0.35$ (from top to
  bottom for $w(y)$ or from left to right for $V$) are shown by dashed
  ones.}
  \label{fig:fig3}
\end{figure}

\section{Reconstructing the Equation of State and the Effective
  Potential from Simulated Data}

We simulate the actual reconstruction procedure using numerically
generated data $r_i=r(y_i)+\delta r_i$ $(i=1,\dots,N)$ with $\delta r_i$ 
being Gaussian distributed (zero mean and variance $\sigma r(y_i)$).
The simulated data assumes a cosmological model with $\Omega_M=0.30$.
We consider $N=30$ data and take $\sigma=0.03$ or $\sigma=0.005$. 
The former error is the distance error of the binned data expected
{}from observations of 200 supernovae by SNAP, while the latter is for
6000 supernovae. We only consider statistical uncertainties.  

We distribute the data uniformly in $y$ from $y=0.95~(z=0.11)$ to
$y=0.60~(z=1.78) $. We perform thousands of Monte-Carlo realizations. 
The 68\% confidence intervals of the reconstructed potentials are shown
in Fig. 4 and Fig. 5. The horizontal axis is the averaged value of
$\hphi$, while the vertical axis is the averaged value of $\hv$. 
The effect of the reduced error may be dramatic. 
We allow for 10\% error in $\Omega_M$ to assess the ambiguity in the
reconstruction of the potential. Obervations of very distant supernovae
at $z\geq 3$ by NGST (Next Generation Space Telescope) and/or
observations of galaxy cluster abundance by SZE (Sunyaev-Zel'dovich
Effect) survey will determine $\Omega_M$ to a few \% \cite{efs}. 

It should be noted that the fitted
$\hr(y)$ does not necessarily satisfy the positivity of the
right-hand-side of Eq.(\ref{reconst:p}), that is, the weak energy
condition. We perform the reconstruction using only the data which
satisfy the weak energy condition.\footnote{Conversely,
  Eq.(\ref{reconst:p}) might provide an upper bound on $\Omega_M$, if we 
  assume that the dark energy respects the weak energy condition,
  although it is not always the case \cite{phantom}.} 
That induces the bias toward larger $\dot \phi^2$. This is why the range 
of $\hphi$ is larger than that of the true one and why the intervals of
the reconstructed potential is distributed downward. Such an effect is 
particularly significant for the cosine type potential because in this
model the equation of state is almost $w=-1$ at  higher redshift. 
Therefore, good estimate of the {\it upper} bound of $\Omega_M$ is
crucial for the success of the reconstruction. 

\begin{figure}[htdp]
  \begin{center}
  \leavevmode\psfig{figure=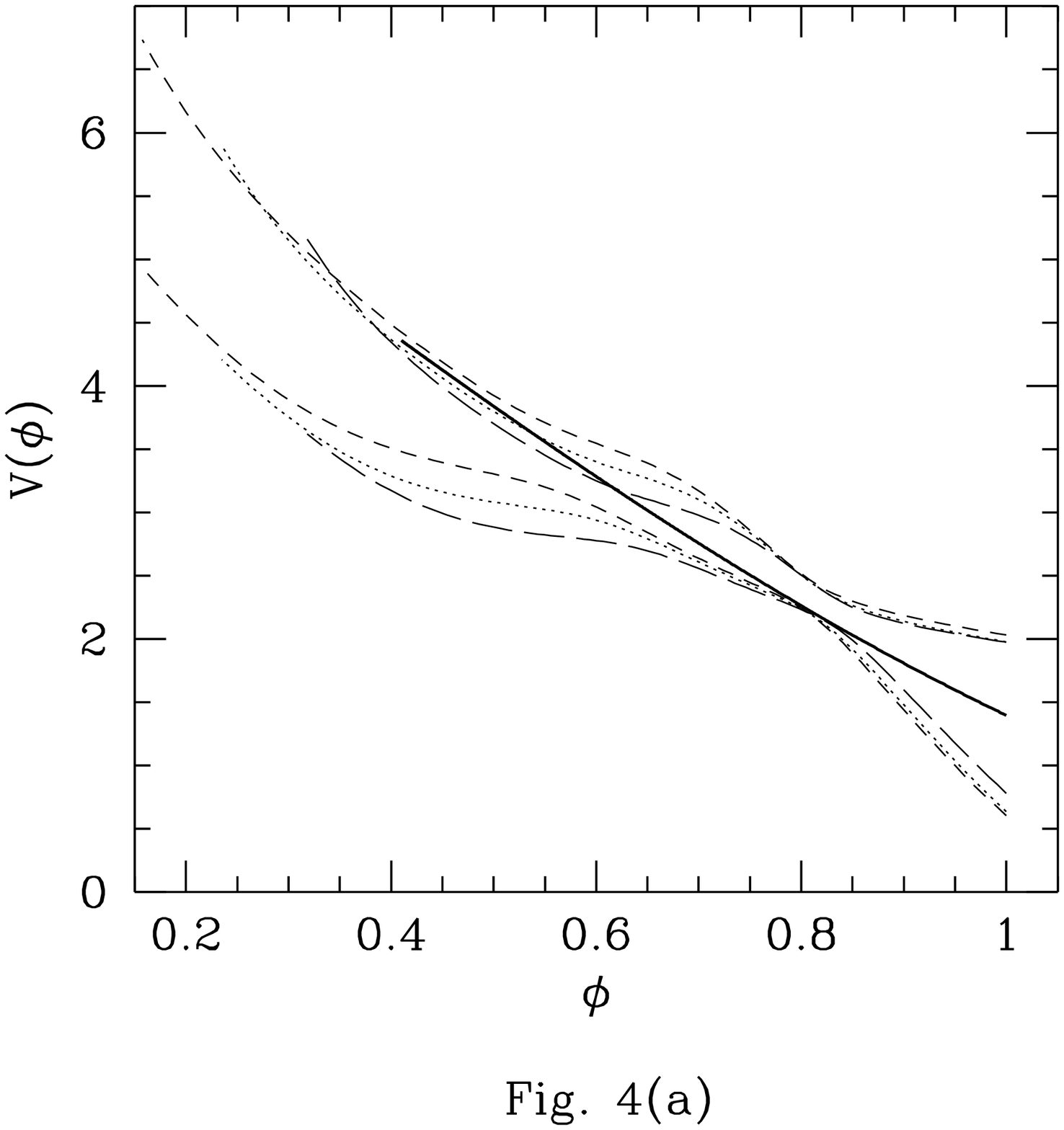,width=8cm}
  \leavevmode\psfig{figure=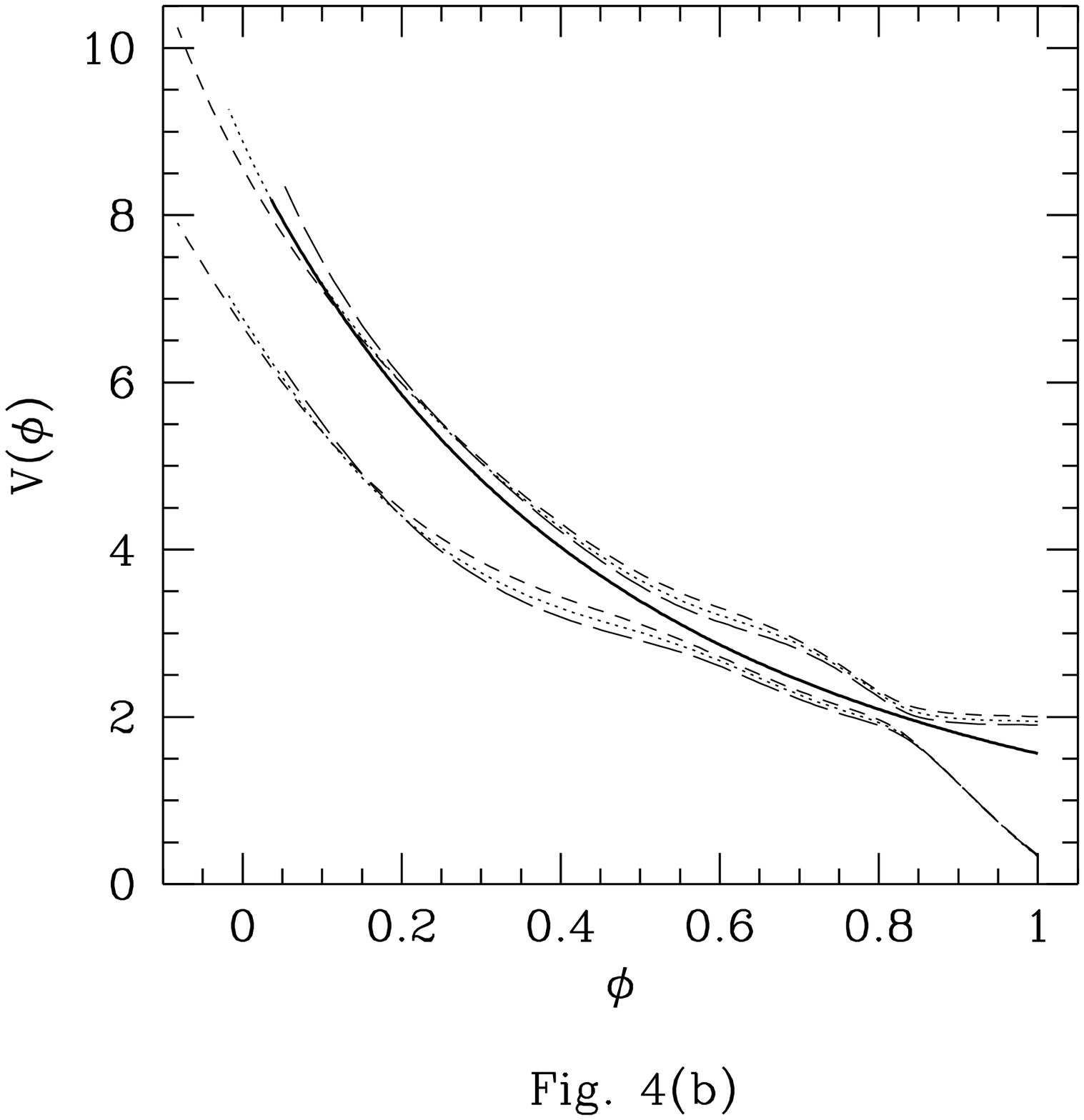,width=8cm}
  \leavevmode\psfig{figure=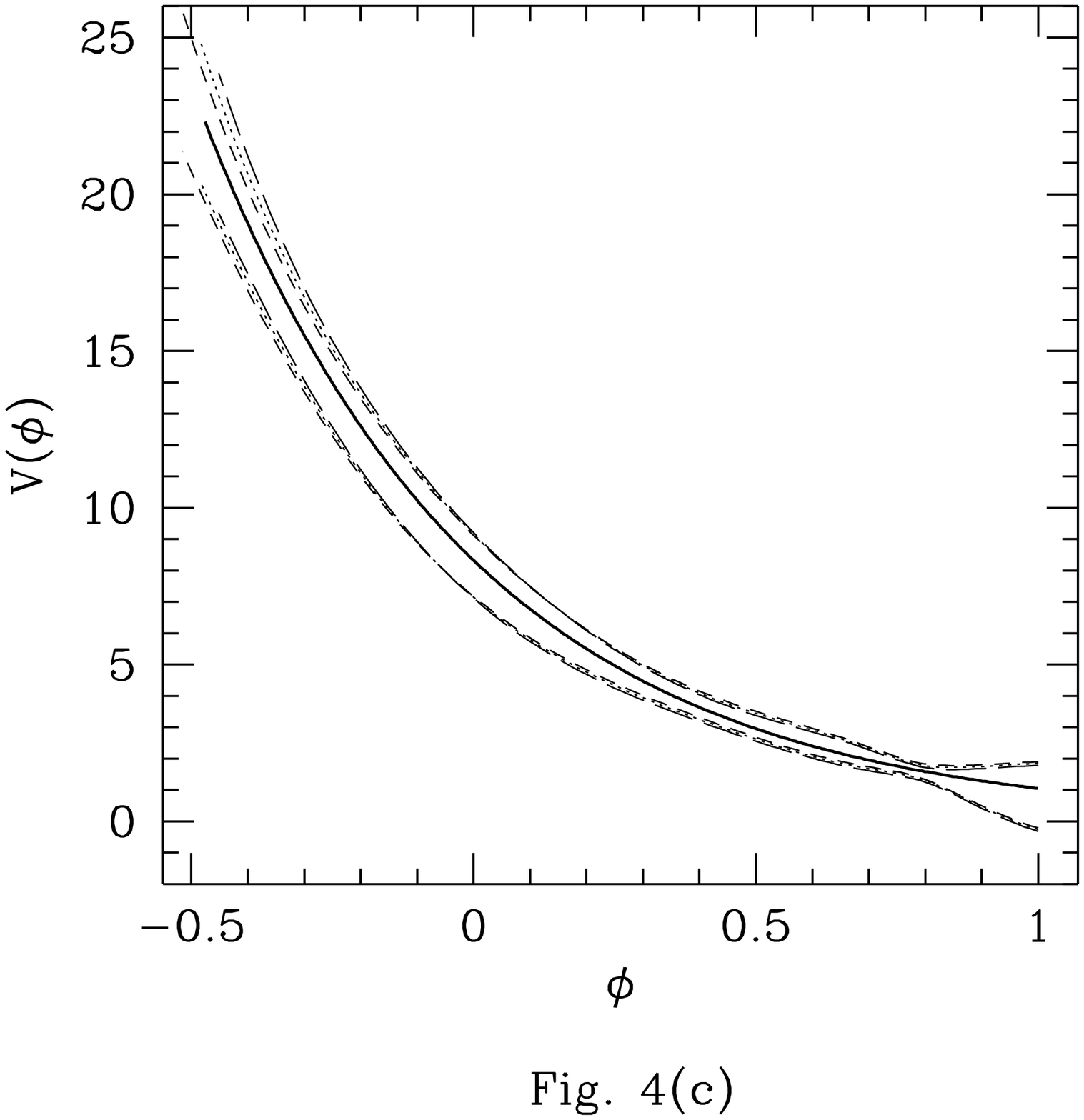,width=8cm}
  \end{center}
  \caption{One sigma intervals for  the reconstructed quintessential 
potential assuming luminosity distance error of 3\% with $N=30$ data: 
(a) cosine potential, (b) inverse power law potential, (c) exponential 
potential. Solid curves are the original potentials. Dotted curves are 
for $\Omega_M=0.30$, short dashed curves for  
$\Omega_M=0.27$, long dashed curves for $\Omega_M=0.33$}
  \label{fig:fig4}
\end{figure}

\begin{figure}[htdp]
  \begin{center}
  \leavevmode\psfig{figure=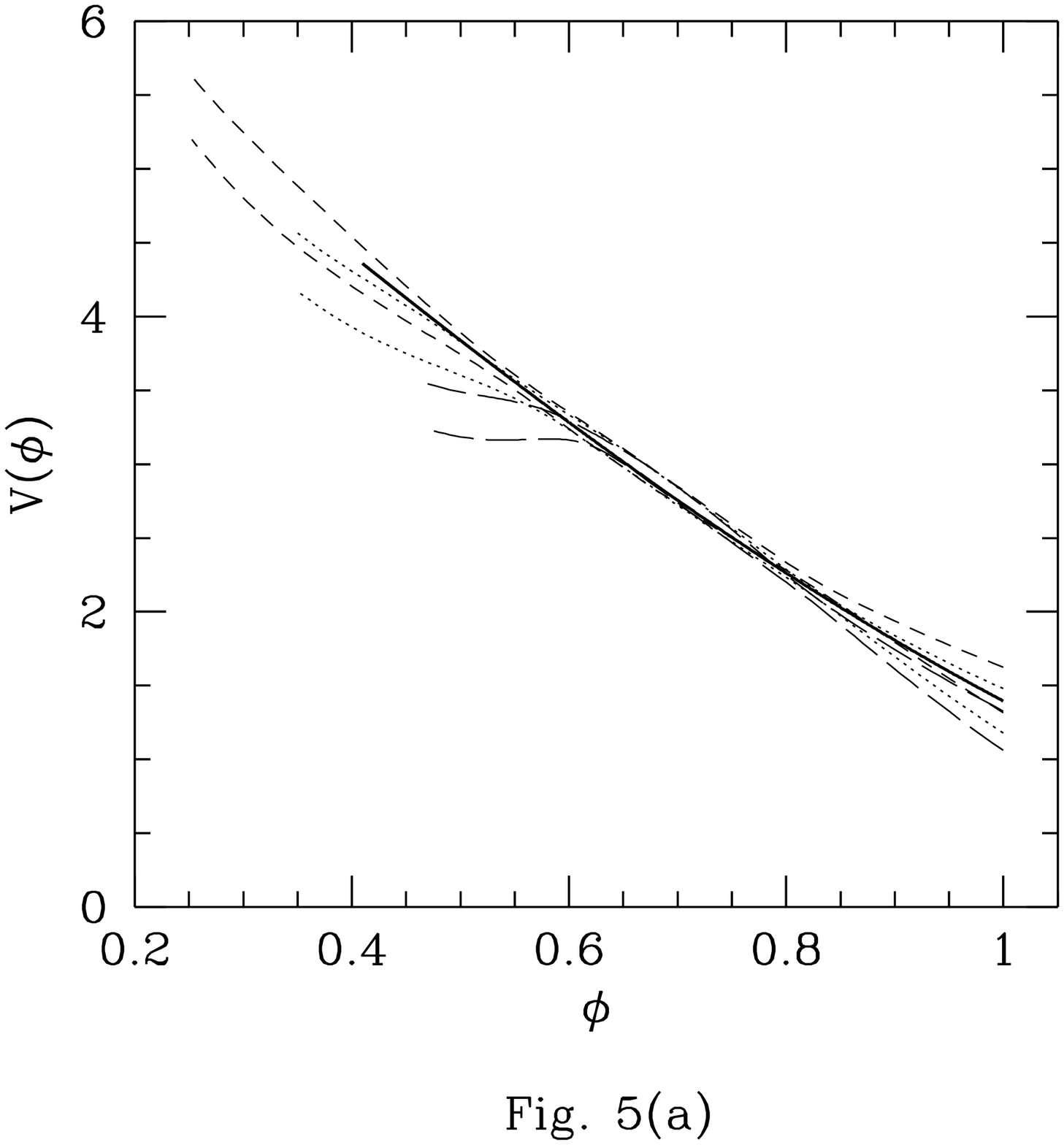,width=8cm}
  \leavevmode\psfig{figure=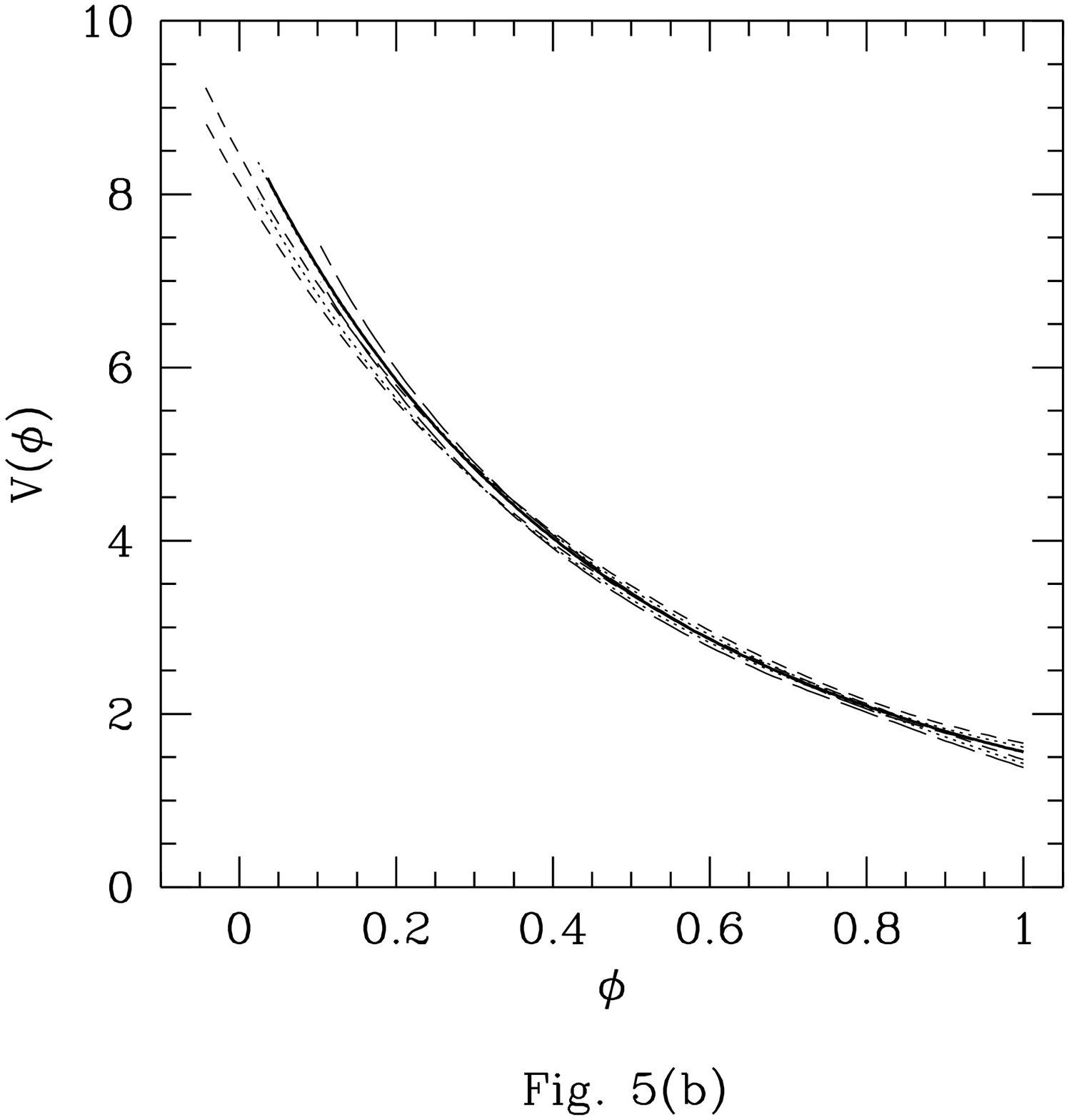,width=8cm}
  \leavevmode\psfig{figure=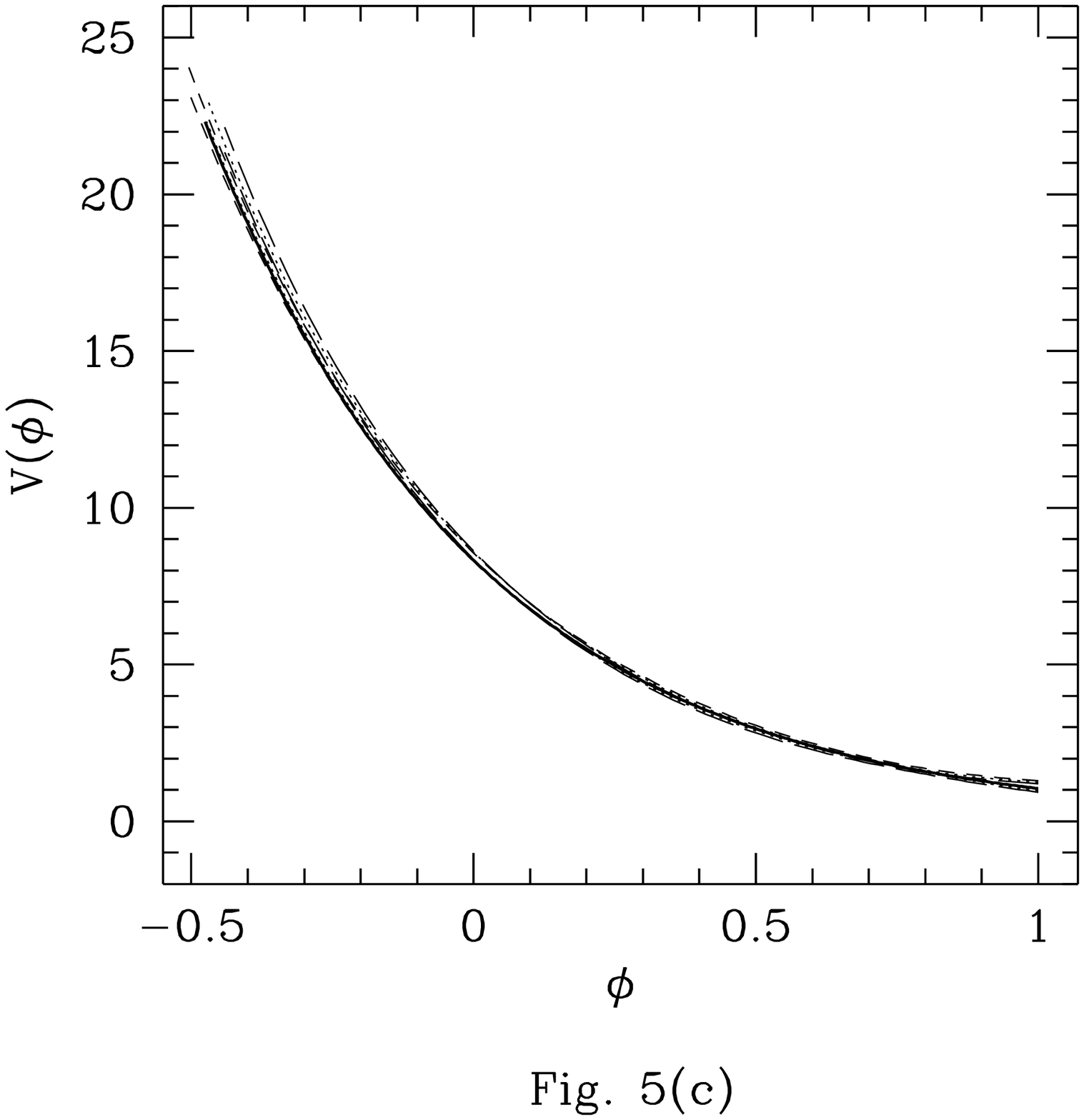,width=8cm}
  \end{center}
  \caption{One sigma intervals for  the reconstructed quintessential 
potential assuming luminosity distance error of 0.5\%. 
The meaning of the curves is the same as Fig.4.}
  \label{fig:fig5}
\end{figure}

\section{Summary}

We have studied the feasibility of reconstructing the equation
of state of dark energy and the effective potential of the quintessence
field from SNIa data by taking into account the uncertainty in
$\Omega_M$ as well as the error in the luminosity distance. 
We have found that $w$ and the range of $\hphi$ are dependent on the 
assumed value of $\Omega_M$, while the whole shape of $\hv(\hphi)$ is less
sensitive to the uncertainty in $\Omega_M$. 
If $\Omega_M$ could be constrained to some 10\% accuracy by other 
observations, which may not be unrealistic expectation \cite{efs}, then 
future high precision measurements of distances to thousands of SNIa
could reveal the shape of the quintessential potential. 


\acknowledgments
One of the authors (TC) would like to thank Misao Sasaki and Fumio
Takahara for useful comments and Dragan Huterer for useful discussion at
the early stage of this work. 
  This work was supported in part by
Grant-in-Aid of Scientific Research of the Ministry of Education,
Culture, and Sports, No.11640274 and 09NP0801.


\end{document}